\documentclass[aps,prb,showpacs,floatfix,superscriptaddress,twocolumn]{revtex4-1}

\pdfoutput=1

\usepackage{natbib} 
\usepackage{amsmath} 
\usepackage{amssymb} 
\usepackage{graphicx}

\usepackage{fancyref}

\begin{document}

\title{Caveats of mean first-passage time methods applied to the crystallization transition: effects of non-Markovianity}

\author{Swetlana Jungblut}
\affiliation{Faculty of Physics, University of Vienna, Boltzmanngasse 5, 1090 Wien, Austria}
\author{Christoph Dellago}
\affiliation{Faculty of Physics, University of Vienna, Boltzmanngasse 5, 1090 Wien, Austria}

\begin{abstract}
Using the crystallization transition in a Lennard-Jones fluid as example, we show 
that mean first-passage time based methods may underestimate the reaction rates. We 
trace the reason of this deficiency back to the non-Markovian character of the 
dynamics caused by the projection to a poorly chosen reaction coordinate. The 
non-Markovianity of the dynamics becomes apparent in the behavior of the recurrence 
times.

\end{abstract}

\pacs{64.70.D-, 64.60.qe, 64.75.Gh}

\maketitle

\section{Introduction \label{intro}}

In the study of activated events such as chemical reactions or first order phase transitions, the calculation of 
rate constants is an important task and, in the past decades, several advanced computational methods have been 
developed for this purpose \cite{dellago:2009,erp:2005,allen:2009,bonella:2012}. These methods concentrate 
on the rare barrier crossing events such that they are not hampered by long waiting times between events. 
However, these techniques are computationally demanding and it may be convenient to switch to 
the mean first-passage time (MFPT) or mean lifetime (MLT) methods, if it is possible 
to observe the reactions on the timescale of straightforward simulations. 
These approaches, based on a statistical description of activated events, have been around for 
years (see, {\it e.g.}, Ref. \citenum{landauer:1989} for 
a historical view), but recently, as simulations became fast enough to 
produce reaction trajectories directly, they have gained popularity \cite{kramers:1940,langer:1969,talkner:1987,haenggi:1990,bicout:1997,rubi:2001,park:2003,berezhkovskii:2005,reguera:2005,bartell:2006,wedekind:2007,wedekind:2008,chkonia:2009,lundrigan:2009,baidakov:2010,baidakov:2011a,baidakov:2011b,baidakov:2012,mokshin:2012,mokshin:2014,shneidman:2014}. Particularly notable 
is the method by Wedekind, Strey and Reguera \cite{wedekind:2007}, which details how reaction rate constants and 
sizes of critical clusters can be extracted from mean first-passage times. 
MFPT and MLT methods have been applied to various processes including the crystallization of a 
Lennard-Jones (LJ) liquid \cite{lundrigan:2009,baidakov:2010,baidakov:2011a,baidakov:2011b,baidakov:2012}. 
Here, we report a 
disagreement that we found between the crystallization rates computed with the MFPT (MLT) method and 
transition interface path sampling \cite{erp:2005} (TIS). This deviation can be traced back to the non-Markovian 
character of the crystallization transition in terms of the chosen reaction coordinate, implying that the 
application of the MFPT and MLT techniques are not as straightforward as suggested in recent studies.

The main issue discussed previously concerning the applicability of MFPT and MLT techniques is the 
lack of timescale separation between 
the nucleation and growth times for processes with relatively low activation 
barriers \cite{wedekind:2007,mokshin:2014,shneidman:2014}. 
In this paper, we point out another source of error, which is related to a 
poor choice of the reaction coordinate. As is known from previous studies 
\cite{tenwolde:1996,moroni:2005,beckham:2011,jungblut:2013a}, 
the crystallization of an undercooled LJ fluid proceeds via the formation of a crystallite with 
body-centered cubic (bcc) structure, which subsequently relaxes into the face-centered cubic 
(fcc) structure. In computational studies of this process one usually defines the number of particles in the 
largest crystalline cluster \cite{steinhardt:1983, tenwolde:1996} as the reaction coordinate. Although this coordinate 
does not contain enough information 
to precisely describe the progress of the reaction, it performs relatively well in comparison to other order 
parameters \cite{beckham:2011} and, in practice, it is the most convenient and widely used one. 
It has been shown before that a poor choice of the reaction coordinate may result in an overestimation 
of the reaction rate \cite{berezhkovskii:2005}. Here, however, we find that an insufficient reaction coordinate 
used in conjunction with MFPT method leads to an underestimation of the rate constant. 

The article is structured as follows. 
We start with the standard analysis of the MFPT and MLT formalism and compare the obtained reaction rates with 
the values from TIS simulations. Then, we take a closer look at the MFPT data for the crystallization of a liquid 
and show that the process we consider is non-Markovian, in contradiction to the main assumption of the analysis. 
We explain this with the 
features of the crystallization transition, particularly with the lack of a good definition of the 
reaction coordinate. To demonstrate the issue, we consider not only the first passage but also 
subsequent passages and show that the behavior of these later passages indicates the onset of 
relaxation. Finally, we argue that the assumption of the relaxation times being negligible on the timescale of 
the reaction does not apply in the case of the crystallization transition.

\section{Simulation details \label{simdetails}}

We simulated a system of $N=6668$ particles interacting via a standard truncated and shifted Lennard-Jones potential. 
The cutoff distance was set to $r_c=2.5$ (in LJ units, which are used throughout the paper). The particles were confined to 
a cubic box with periodic boundary conditions in all directions, which was allowed to fluctuate to fix the pressure at 
a value close to zero ($p=0.003257$). 
The evolution of the system was simulated with molecular dynamics (MD) simulation in the $NpH$ ensemble \cite{andersen:1980} with 
a time step of $\Delta t=0.01$ and an enthalpy of $H=-5.11$ per particle such that the initial 
undercooling was about $28\%$ ($T=0.5$).
We used Steinhardt bond order parameters \cite{steinhardt:1983} with the standard 
scheme proposed by ten Wolde, Ruiz-Montero, and Frenkel \cite{tenwolde:1996} to identify crystalline 
clusters, and monitored the progress of the reaction by considering the size of the largest crystalline cluster, $n_s$. 
Details of this analysis and of the TIS simulations \cite{erp:2003,erp:2005,moroni:2005a} are extensively described 
in our previous work \cite{jungblut:2011}. Mean first-passage times were 
calculated from a collection of $200$ MD trajectories started in the initial undercooled state. For every 
cluster size considered, we calculated the average time the system needs to form clusters of this size or larger 
for the first time. 
Also, for every trajectory, we computed the lifetime of the undercooled state, which was defined as the 
time to reach a particular cluster size, well above the critical size. 
For this purpose, we chose the size of $n_s=400$ and compared the resulting rate with the one obtained for $n_s=1000$. 
  
\section{Results and discussion  \label{results}}

In this section, we present the reaction rates computed with 
different methods and demonstrate that the crystallization of an undercooled LJ fluid does 
not comply with the requirement for a Markovian process, which the starting point of the 
MFPT analysis. We attribute this behavior to the fact that the nature of the 
crystallization transition is not completely captured by the size of the largest crystalline cluster 
used as reaction coordinate.   

The main assumption of the MFPT and MLT methods is that the dynamics of the reaction in terms of the reaction 
coordinate, $n_s$, is described 
by the one-dimensional Fokker-Planck equation \cite{reguera:2005} for the probability density function $p(n_s, t)$ 
of $n_s$ at time $t$:
\begin{eqnarray}
\frac{\partial p(n_s, t)}{\partial t}&=&\frac{\partial}{\partial n_s} \left \{ D_o e^{-\beta U(n_s)} \frac{\partial}{\partial n_s}\left [ p(n_s,t) e^{\beta U(n_s)} \right ]\right \} \nonumber \\ 
&=&-\frac{\partial j}{\partial n_s}, \label{fpe}
\end{eqnarray}
where $j$ is the probability flux, $D_0$ is the diffusion coefficient (assumed to be constant here), 
$U$ is the free energy, $T$ is the temperature, $k_B$ is the Boltzmann constant, and $\beta=1/k_BT$. 
For an activated process with a relatively high nucleation barrier, one 
assumes that the system rapidly reaches a steady-state with a constant probability current 
\begin{equation}
j=-D_o e^{-\beta U(n_s)} \frac{\partial}{\partial n_s}\left [ p_{\rm st}(n_s) e^{\beta U(n_s)} \right ]. \label{fpeStat}
\end{equation}
Here,  $p_{\rm st}(n_s)$ is the stationary distribution of states, and the reaction rate $j$ is related to 
the nucleation rate $J$ by $j=JV$. 

\subsection{Reaction rates}
\subsubsection{Mean first-passage time}

From Eq.~\ref{fpeStat}, the MFPT \cite{moss:1989,haenggi:1990} for a state with a given number of particles in the 
largest 
crystalline cluster, $n_s$, can be calculated as  
\begin{equation}
\tau(n_s)=\frac{1}{D_0}\int_{n_0}^{n_s} dy \exp \left[ \beta U \left(y\right) \right ] \int_{a}^y dz \exp \left[ -\beta U \left(z\right) \right ], \label{mfptW}
\end{equation}
where $a$ is the 
reflective boundary of the initial state and $n_0$, from which the times to reach a particular cluster size are calculated, belongs to the 
metastable state. 
For the case of the crystallization transition, we set $a=0$ (no crystallites) and $n_0=20$. In fact, for $n_0$ any value smaller than 
the position of the top of the free energy barrier is valid and we select this one following our definition of the 
initial state which we used in TIS simulations.   

In the scheme proposed by Wedekind, Strey and Reguera \cite{wedekind:2007}, the behavior of the MFPT close to the transition region 
is described by the function 
\begin{equation}
\tau(n_s)=\frac{\tau_J}{2}\left \{ 1 + {\rm erf} \left (\left[n_s-n_s^*\right]c\right)\right\}, \label{mfptErf}
\end{equation} 
where ${\rm erf}(x)=(2/\sqrt{\pi})\int_0^x\exp(-y^2)dy$ is the error function, $n_s^*$ is the size of the critical cluster, and 
$c=\sqrt{0.5\beta \left| U^{''}(n_s^*)\right|}$ is the local curvature at the top of the barrier. The 
reaction rate is then given by 
\begin{equation}
JV=\frac{1}{\tau_J}, \label{nuclrate}
\end{equation} 
where $V$ is the volume of the system.
\begin{figure}[tb]
\begin{center}
\includegraphics[clip=,width=0.95\columnwidth]{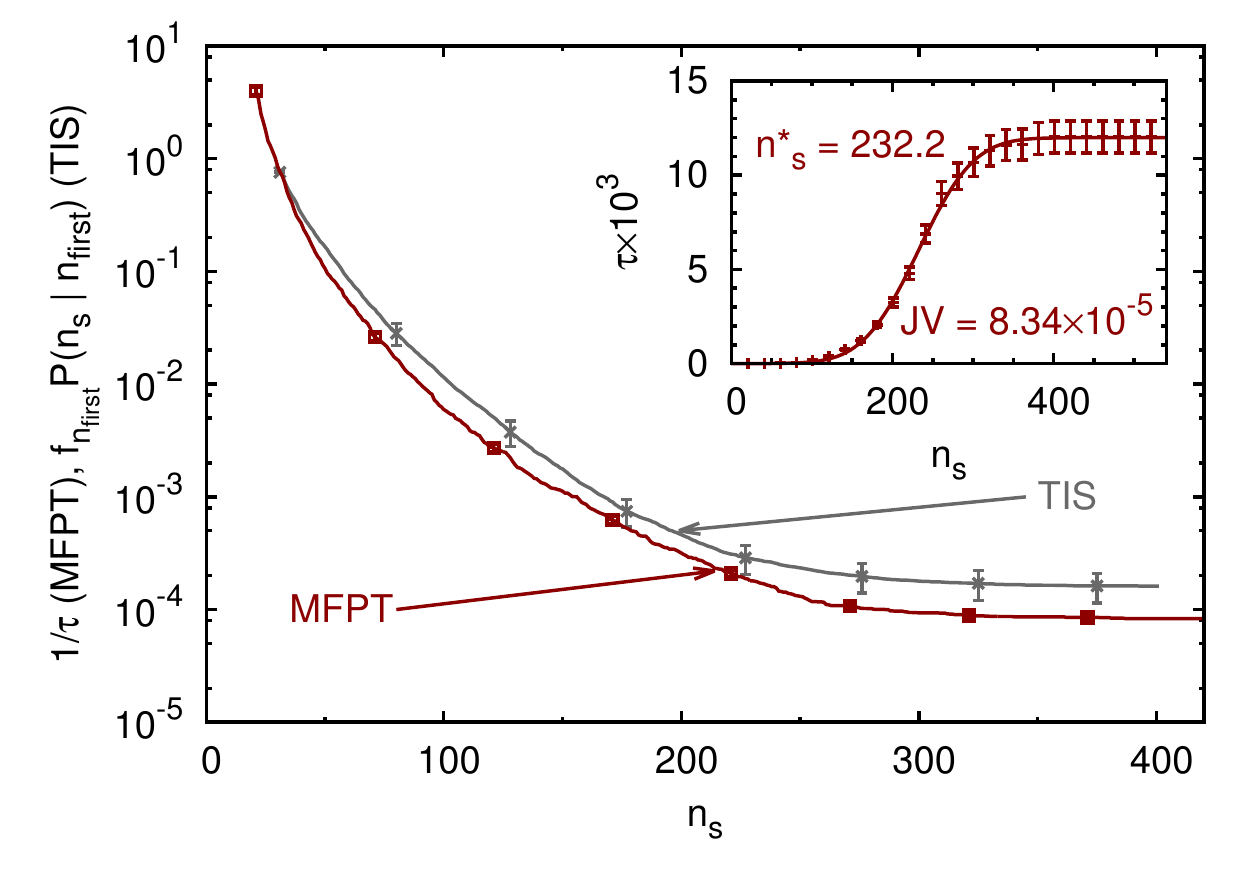} 
\caption{\label{rates} The inverse of the MFPT, $1/\tau$, and the product of the flux out of the initial state and the 
TIS conditional probability to reach a given state, $f_{\rm first}P(n_{s}|n_{\rm first})$, 
as a function of the cluster size, $n_s$. 
For large cluster sizes, both curves saturate to the values of the reaction rate, $JV$. 
For clarity, errors are indicated only for selected data points.   
Inset: Selected values of MFPTs with errors and the fit of all data points to Eq.~\ref{mfptErf} (solid line). Also included are 
the values of fitted parameters $JV$ and $n_s^*$. 
} 
\end{center}
\end{figure}

As can be seen in the inset of Fig.~\ref{rates}, Eq.~\ref{mfptErf} perfectly reproduces the MFPTs obtained 
in the simulations if $\tau_J$, $n_s^*$, and $c$ are used as fitting parameters. The fit yields a critical cluster size 
of $n_s^*=232$, a crystal nucleation rate of $JV=8.3\times10^{-5}$, and a local curvature at the top of the barrier of $c=0.013$. 
The timescales of the nucleation and growth regimes 
are clearly separated, as can be seen from the shape of the MFPT curve, which is displayed in the inset of 
Fig.~\ref{rates}. Also, in section \ref{tis}, we present the lengths of the 
crystallizing paths, which go directly from the metastable to the crystalline states. These times 
the system needs to grow crystalline clusters are distinctly shorter than the MFPTs.    

In Fig.~\ref{rates}, we plotted the inverse of the MFPTs as a function of the cluster size in comparison 
with the results of the TIS simulations discussed below.   
One can clearly see that the MFPT method underestimates the reaction rate obtained with TIS 
by almost a factor of two ($1.92$). 

\subsubsection{Mean lifetime} \label{mlt}

\begin{figure}[tb]
\begin{center}
\includegraphics[clip=,width=0.95\columnwidth]{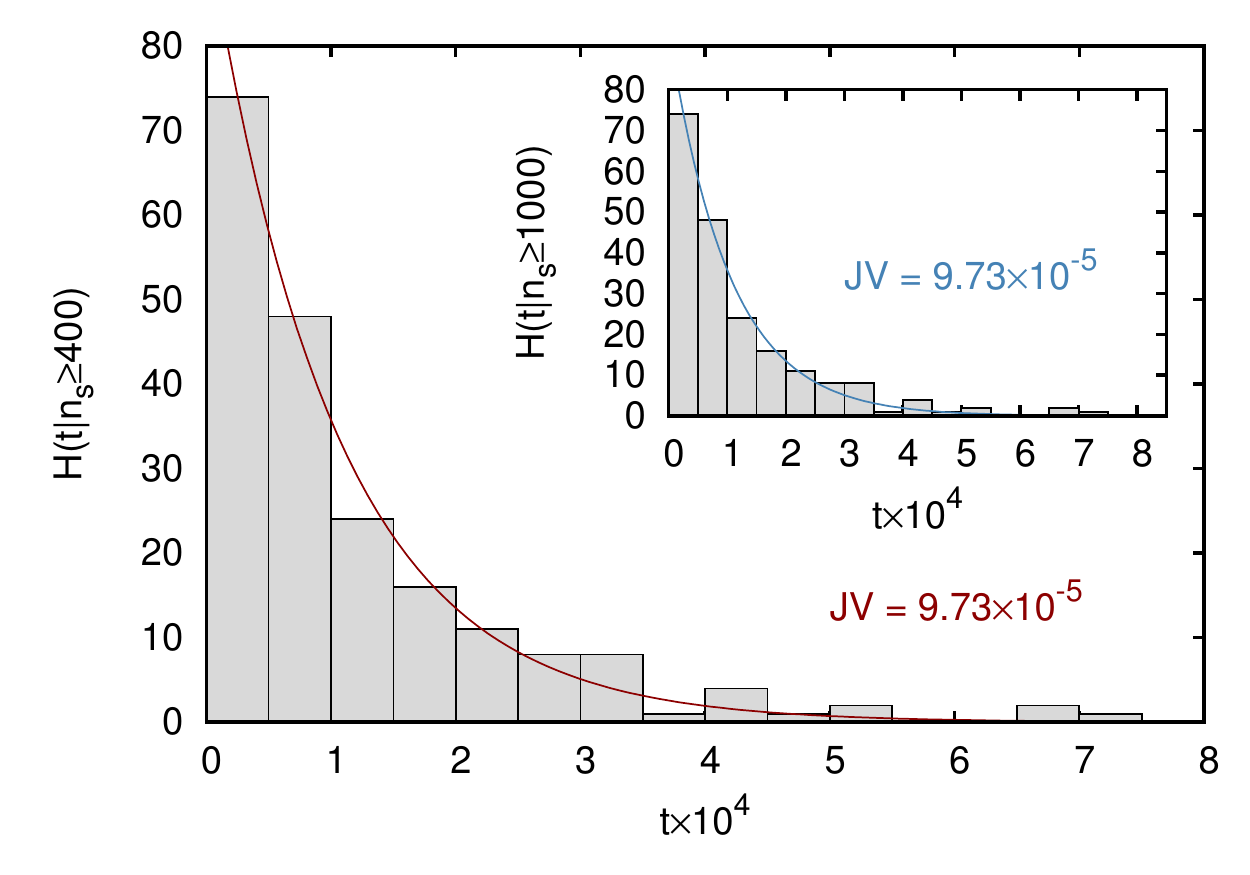} 
\caption{\label{meanlifetime} Histogram of the lifetimes of the metastable state defined as $t|n_s\geq400$ (main frame) 
and $t|n_s\geq1000$ (inset) and fits to Eq.~\ref{distribution}. Also included are 
the values of the fitted parameter $JV$. } 
\end{center}
\end{figure}

The mean lifetime (MLT) \cite{baidakov:2011a} or direct observation method \cite{chkonia:2009} is 
based on the same formalism as the MFPT method, and it is expected to give the same results if the final states are well beyond 
the critical region \cite{boilley:2004}. 
It also does not rely on the exact definition of the transition state 
\cite{talkner:1987} and allows a distinction between nucleation and growth regimes \cite{shneidman:2014}.
In Fig.~\ref{meanlifetime}, we present the probability to observe 
a crystallized system in a given time interval, which is fitted 
to the predicted Poisson distribution \cite{baidakov:2011a} 
\begin{equation}
H(t|n_s\geq400)=h t \exp (-\lambda t), \label{distribution}
\end{equation}
using $h$ and $\lambda$ as fitting parameters. The reaction rate $JV$ is 
then equal to  $\lambda$: 
\begin{equation}
JV=\lambda. \label{rateMLT}
\end{equation}
As expected, the rate we obtain with the MLT method ($JV=9.7\times10^{-5}$) is comparable to the MFPT rate.   
The inset of Fig.~\ref{meanlifetime} also demonstrates that if we choose another value to define a 
crystallized system ($n_s=1000$) the reaction rate does not change, provided we stay well above the 
critical cluster size. 
Thus, similarly to the MFPTs, the contribution of the growth times is negligible.  
Still, the value of the reaction rate does not coincide with the one obtained with the TIS method. 

\subsubsection{Transition interface sampling} \label{tis}

The advantage of the transition interface sampling technique is that it does not depend on the 
definition of the reaction coordinate. One only has to be able to distinguish between the initial and the final 
states of the reaction and sample paths in the transition region between these states. 
In the TIS formalism, the crystallization rate is expressed as 
\begin{equation}
JV=f_{\rm first}P(n_{\rm last}|n_{\rm first}), \label{rateTIS}
\end{equation}
where $f_{\rm first}$ is the flux through the first interface 
considered for sampling, $n_{\rm first}=30$, and $P(n_{\rm last}|n_{\rm first})$ is the probability 
to reach the final state under the condition that the trajectory, which started in the 
initial phase ($n_{0}\leq20$), crossed the first interface ($n_{\rm first}=30$).  
The definition of the final state is relatively uncritical, since the probability to 
relax into the final state becomes constant after the system overcomes the free energy barrier. 
The value of the reaction rate we obtained in the TIS simulations \cite{jungblut:2011} is 
$JV=(1.6\pm0.4)\times10^{-4}$, which differs significantly from the rates obtained with the MFPT 
and MLT methods.  

\begin{figure}[tb]
\begin{center}
\includegraphics[clip=,width=0.95\columnwidth]{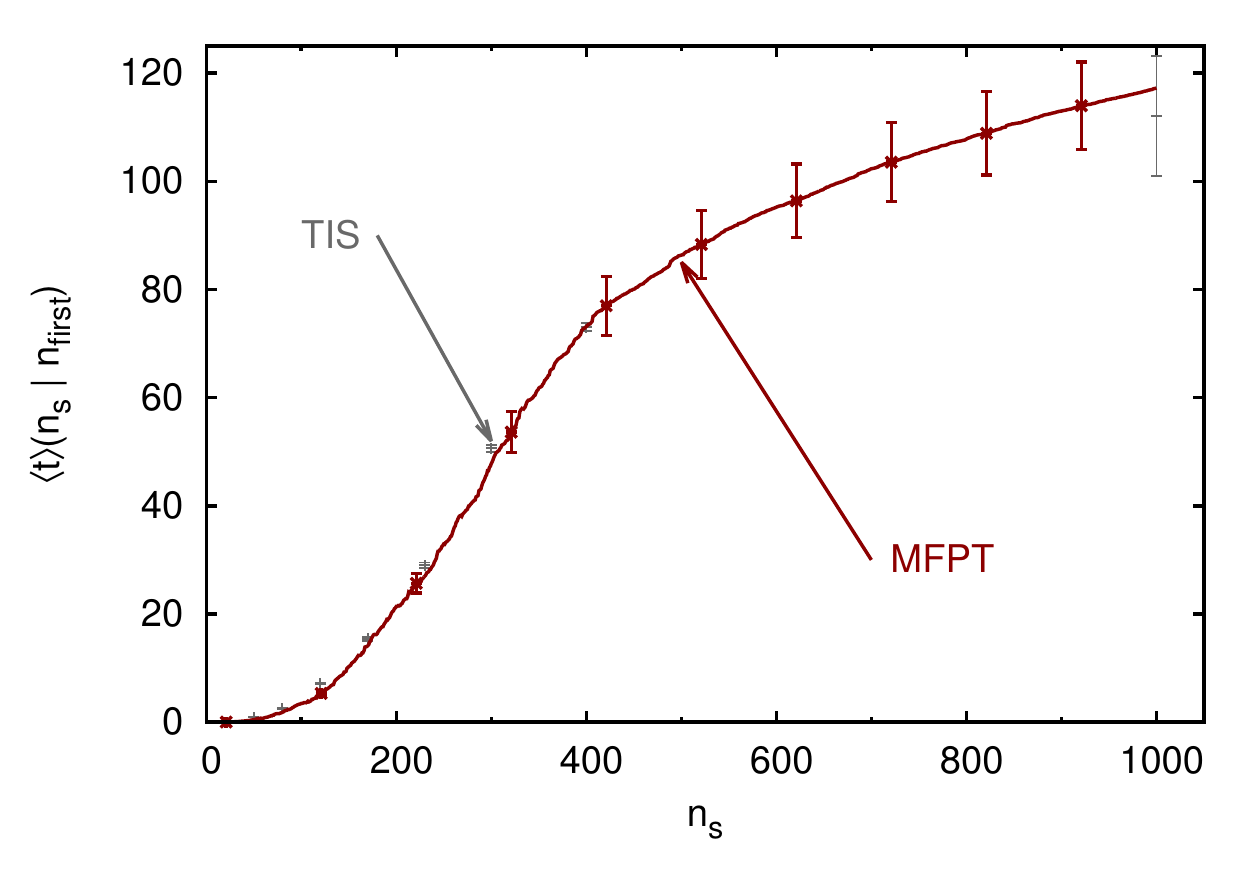} 
\caption{\label{meangrowthtime} Mean times to reach a particular cluster size for the first time when coming 
directly from the border of the initial state ($n_s=20$). Lengths of TIS paths are evaluated at the positions of the 
interfaces used to calculate the crystal nucleation rate ($n_s=50, 80, 120, 170, 230, 300, 400$), the last 
point ($n_s=1000$) resulted from average over $100$ 
crystallizing paths. MFPTs are evaluated at all cluster sizes, 
while errors are presented only for selected values. } 
\end{center}
\end{figure}

Still, the paths which are sampled in the TIS are essentially the same trajectories as those 
considered for the MFPT calculations. For example, in Fig.~\ref{meangrowthtime}, we compare 
the lengths of the TIS and MFPT 
paths directly connecting the initial state with the respective states with a given cluster size. The 
length of the MFPT trajectories is restricted to the fragments of the trajectories, in which the system leaves 
the initial state $n_0\leq20$ for the last time and reaches the given state. 
Evidently, there is no difference in lengths between the growing paths.  
Also the flux from the initial state 
used in TIS corresponds to the inverse MFPT at the position of the first TIS interface, as can be seen in 
Fig.~\ref{rates}, where these 
values coincide since the probability $P(n_{\rm first}|n_{\rm first})$ equals unity. 

In addition, we performed a commitment analysis to find the transition states of the system \cite{jungblut:2011a}. The 
configurations with equal probabilities to reach either the initial undercooled liquid or the final 
fully crystalline states contain crystalline clusters with sizes between $118$ and $263$ particles. 
The critical cluster size obtained with the MFPT method lies well within this range. 

\subsubsection{Multiple crossings} \label{mc}

\begin{figure}[tb]
\begin{center}
\includegraphics[clip=,width=0.95\columnwidth]{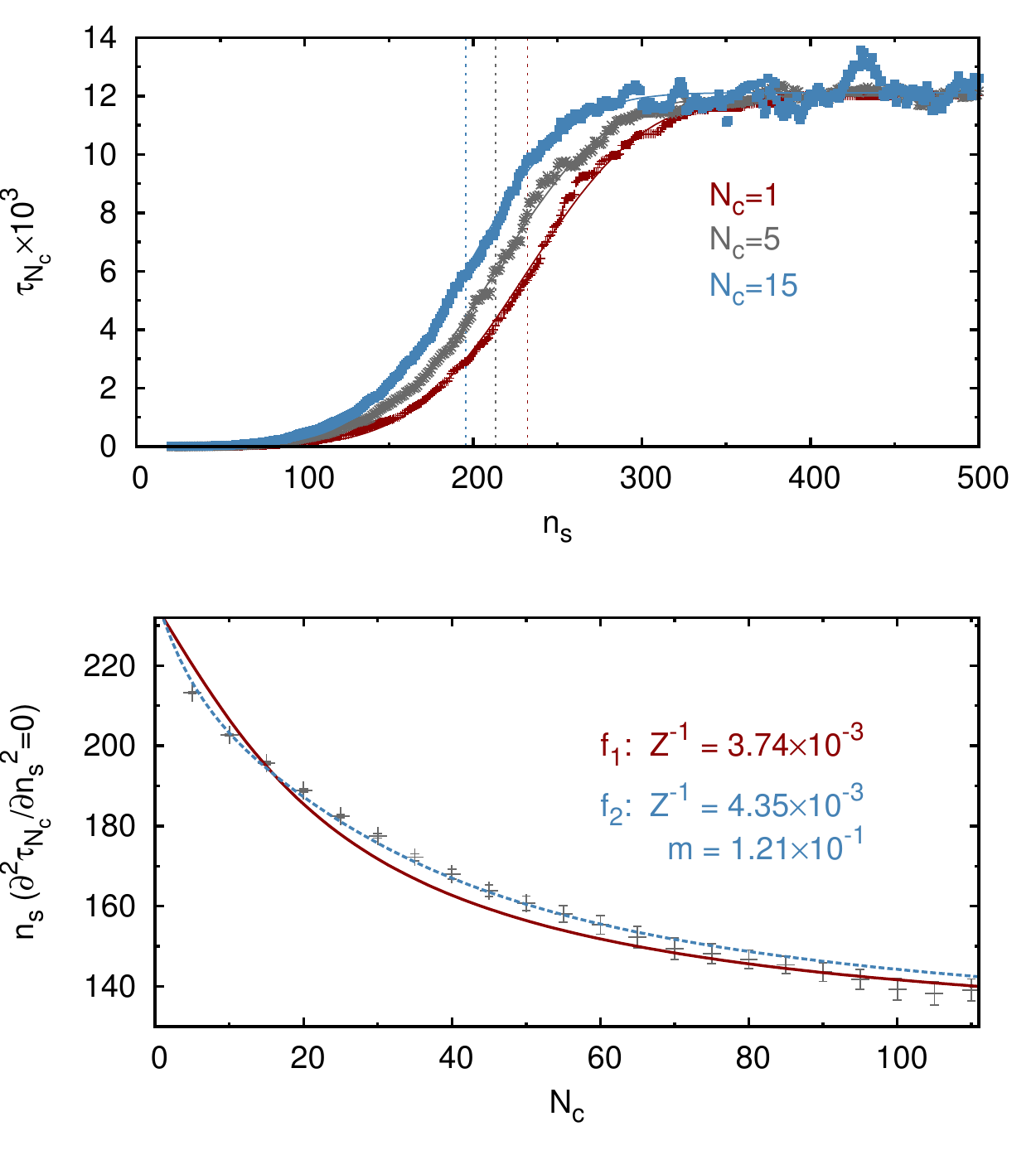} 
\caption{\label{fitsMC} Top: Mean times for $1^{\textrm st}$, $5^{\textrm th}$ and  $15^{\textrm th}$ passages at 
given cluster size, $n_s$. Solid lines indicate fits to Eq.~\ref{mfptErf}, from which we extract the positions of 
the inflection points. Bottom: 
Position of the inflection point as a function of the number of crossings, $N_c$, (values are extracted from 
corresponding fits and error bars indicate the error of the fits only). Lines are fits to Eq.~\ref{inflpointposition} 
with $f_1(N_c)=N_c-1$ (solid line, fitting constant $Z$) and $f_2(N_c)=(N_c-1)\{1+\exp(-m[N_c-1])\}$ 
(broken line, fitting constants $Z$ and $m$). } 
\end{center}
\end{figure}

Next, we looked at the times of subsequent passages at a given cluster size. 
All clusters considered here have a finite probability to shrink 
to smaller sizes, which depends on the size of the cluster. Hence, particularly smaller clusters tend 
to fluctuate around a size region for a while, passing through an imaginary interface multiple times. 
We thus collected the times a certain cluster size is reached from below for the first (MFPT), second, and 
further times.      
In the top panel of Fig.~\ref{fitsMC}, we plotted the corresponding averages over all trajectories for selected passages. 
The data for larger cluster sizes becomes increasingly noisy with the order of passage, 
since the number of trajectories, in which multiple crossings are observed, decreases.
The shape of the obtained curves for subsequent crossings is similar to that of the MFPT, 
but the position of the inflection point varies with the number of 
passages. This behavior is expected since the integration of the process in MD is discrete in time, 
and the recurrence times for a state depend on the time interval $\Delta t$ between configurations \cite{smoluchowski:1915,kac:1947,balakrishnan:2000}. 
In a Markovian process, described in the framework of the Fokker-Planck formalism, the  
position of the inflection point for the $N_c^{\rm th}$ passage can be approximated as (see Appendix~\ref{appendix})    
\begin{equation}
\tilde{n}_s(N_c)=n_s^*+\frac{\sqrt{\pi}Z^{-1}}{2c^2\Delta t f(N_c)}\left \{ 1 \pm \sqrt{1+\frac{8 \left [c\Delta t f(N_c)\right ]^2}{\pi Z^{-2}}}\right \},\label{inflpointposition}
\end{equation} 
with $f(N_c)=N_c-1$, $c$ and $n_s^*$ extracted from the fit of Eq.~\ref{mfptErf} to the MFPTs. 
$Z$ is the normalization constant of the probability distribution 
which we use as a fitting parameter. In Fig.~\ref{fitsMC} (bottom panel), we demonstrate that the values for the position of the inflection points are relatively close to 
this function. However, the shape of the curve can be reproduced almost exactly by using $f(N_c)=(N_c-1)\{1+\exp(-m[N_c-1])\}$ 
with another fitting constant $m$. We motivate this choice with the decreasing variations of the recurrence times, which is 
a feature of a non-Markovian process. 
According to the fit shown in Fig.~\ref{fitsMC}, $m\approx0.1$ indicates that the memory effects have effectively 
decayed after about $10$ recurrences, corresponding to a decay time of $\sim 10^3$ at the top of the barrier, which is 
comparable to the MFPTs. This is in accordance with the observed effect on the crystallization rate, since a shorter 
relaxation time would not influence the MFPT, while a longer relaxation would produce a larger discrepancy between the 
reaction rates computed with the MFPT and TIS methods.      

\subsubsection{Mean recurrence times} \label{mrt}

\begin{figure}[tb]
\begin{center}
\includegraphics[clip=,width=0.95\columnwidth]{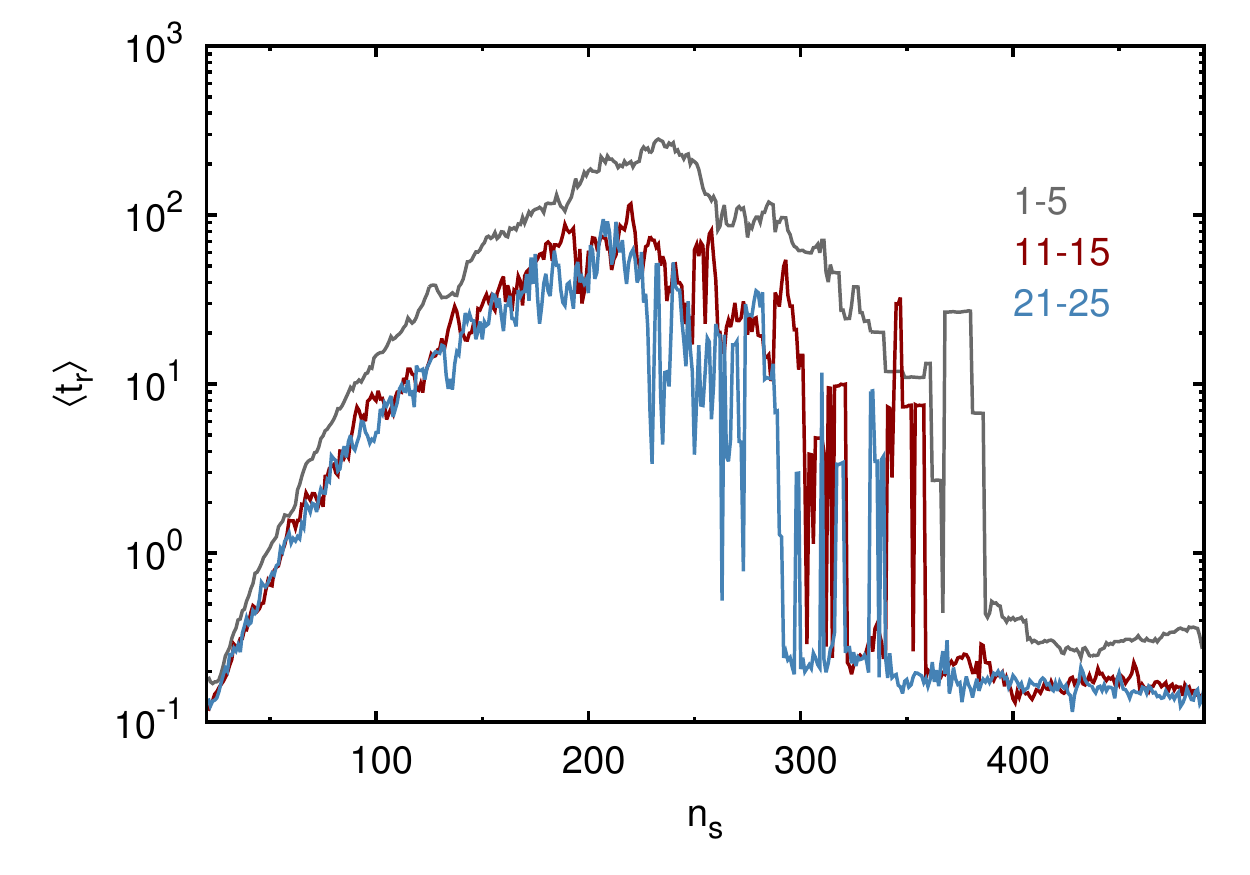} 
\caption{\label{recurrence} Mean recurrence times for given cluster sizes calculated as half of the 
mean times between subsequent passages and averaged over $1^{\textrm st}$ to $5^{\textrm th}$,  
$11^{\textrm th}$ to $15^{\textrm th}$, and $21^{\textrm st}$ to $25^{\textrm th}$ passages.} 
\end{center}
\end{figure}

In the period between two subsequent passages at a given size, the number of particles in the crystalline cluster is first 
above and then below this value. 
Generally, the time to shrink to a given cluster size differs from the time to 
grow to this size. Thus, we define a mean recurrence time as half of the time between subsequent crossings of 
an imaginary interface, at which the cluster is growing. 
To improve the statistics, we also average the obtained values over 
five subsequent passages. In Fig.~\ref{recurrence}, we thus present the data averaged over ten recurrences. 
Although still quite noisy, the mean recurrence times for the first 
passages are distinctly larger than the times for the following passages, if crystalline clusters are large enough. 
The difference becomes smaller as the order of passages considered increases.  
For a stationary distribution of states, the mean recurrence times are inversely proportional to the 
steady-state probabilities, $p_{\rm st}(n_s)$ \cite{smoluchowski:1915,kac:1947}, 
from Eq.~\ref{fpeStat}: 
\begin{equation}
\langle t_{r} \rangle (n_s)=\frac{\Delta t}{p_{\rm st}(n_s)}.\label{meanrecurrencetime}
\end{equation}   
Here, $\Delta t$ is the time interval between configurations, {\it i.e.}, the MD integration step. 
As Fig.~\ref{recurrence}  demonstrates, in our case, the mean recurrence times vary with the order of visits to a given state.   
This finding contradicts the Markovianity assumed in the MFPT and MLT analysis, which implies that the 
time needed to return to a certain point should only depend on the point but not on the number of 
times the point has been reached before. 
However, after a few passages, the recurrence times become almost constant, indicating that the memory effects decline rapidly, 
as we have assumed in the analysis of the inflection points. 

The memory effect observed in the recurrence times is most likely due to a structural relaxation not described by the 
reaction coordinate. 
The importance of the structure of the crystalline 
clusters has been pointed out in several works \cite{tenwolde:1996,moroni:2005,beckham:2011,jungblut:2013a}, 
which indicate that the crystallization transition in undercooled LJ fluids follows the Ostwald's step 
rule \cite{ostwald:1897}. According to this rule, a metastable system does not have to relax directly into the 
most stable state. If there is another metastable state, the probability that 
the system will visit this state 
depends on the height of the free energy barrier between the states \cite{stranski:1933}. 
For an undercooled LJ system, the most stable state is the face-centered cubic (fcc) structure, but the 
free energy barrier between the liquid and the metastable body-centered cubic (bcc) crystal is 
sufficiently low, such that the fluid may first freeze into the bcc structure and then relax into the more 
stable fcc structure.      

The main message of this scenario is that there 
is a second important variable in addition to the cluster size, which is needed to completely describe the 
transition. However, we cannot deduce a precise definition for this coordinate, only that it is somehow connected 
to the structure of the cluster. We assume that the equilibration of the system along this coordinate 
requires a noticeable amount of time, in which the cluster does not grow further but is also not driven back to 
the initial state. As a consequence, although the trajectory is evolving, the overall reaction rate is not influenced. 
One could try to eliminate 
the non-Markovianity by using an appropriate two-dimensional reaction coordinate \cite{risken:1989,berezhkovskii:2005,peters:2013}, 
but it is unclear how to do this in practice. Another approach would be to keep a one-dimensional reaction coordinate and 
apply more sophisticated models for barrier crossing events that include memory effects, as has been done 
recently for the case of polymer dynamics \cite{makarov:2013}.  

We were particularly surprised to find discrepancies in the results obtained with different methods, because 
the applicability of the MFPT and MLT methods to crystallization transition has been demonstrated in several previous 
works \cite{lundrigan:2009,baidakov:2010,baidakov:2011a,baidakov:2011b,baidakov:2012}. 
We presume that 
the differences we see are due to the relative importance of the structural transition for our system. 
We guess that the non-zero pressure used by other authors suppresses the impact of the 
structural relaxation. In other words, at zero pressure, the timescale for this relaxation becomes 
noticeable in comparison to the timescale of crystal nucleation.  

\section{Summary} \label{sum}

We have studied the kinetics of the crystallization transition with different methods and have demonstrated 
that the quality of the reaction coordinate is important for the application of the mean first-passage 
time techniques. 
While previous studies placed emphasis on the importance of the timescale separation between the nucleation and 
growth processes, we have shown that the nature of the transition process also plays 
an important role. It has been shown before that the results of the MFPT calculations heavily rely on the 
definition of a good reaction coordinate \cite{berezhkovskii:2005}, and here we demonstrate that, 
at least for two-step nucleation processes like crystallization, apparently 
sound results of the MFPT analysis may be wrong. The examination of the subsequent passages 
shed light on the origin of the failure of the MFPT method, indicating that a structural 
relaxation of the crystalline nucleus not captured by the reaction coordinate plays an 
important role.  

In general, the analysis of the mean recurrence times is a straightforward method to 
detect a non-Markovian character of the process in a given reaction coordinate. Thus, 
this approach provides a method to evaluate the reliability of MFPT-based techniques.
It is valid for any kind of transition and, should the recurrence 
times demonstrate appearance of the memory effects, indicates the need for a more 
detailed analysis. Then, one can either try to re-define the reaction coordinate and 
the Fokker-Planck equation, or just use advanced simulation techniques, like TIS, which 
do not rely on a valid reaction coordinate.

\acknowledgments
The computational results presented have been achieved, in part, using 
the Vienna Scientific Cluster (VSC). We acknowledge financial support of the 
Austrian Science Fund (FWF) within the Project V 305-N27 as well as SFB ViCoM (Grant F41). 

\appendix* 
\section{Position of the inflection point for the $N_c^{\rm th}$ passage time} \label{appendix}

We calculate the position of the inflection point for the $N_c^{\rm th}$ passage time in the 
steepest descent approximation used in the MFPT method \cite{wedekind:2007,wedekind:2008}.
The time to reach a given cluster size for the $N_c^{\rm th}$ time consists of the time 
to reach it for the first time and $N_c-1$ returns to this state:  
\begin{equation}
\tau_{N_c}(n_s)=\tau(n_s)+f(N_c)\langle t_r \rangle(n_s), \label{nthtime}
\end{equation}
where $\tau$ is the MFPT (Eq.~\ref{mfptErf}), $\langle t_r \rangle$ is the average recurrence 
time. Evidently, $f(N_c)=(N_c-1)$ for constant recurrence times, which is a signature of a Markovian 
process. In our case, however, the system is better described by $f(N_c)=(N_c-1)\{1+\exp(-m[N_c-1])\}$, 
where $m$ is a fitting constant we introduce to include memory effects.   
In the continuous time limit \cite{balakrishnan:2000}, the recurrence time is infinitely small, even in the 
vicinity of the barrier. However, we consider a discrete time process \cite{smoluchowski:1915,kac:1947} for which this 
time is inversely proportional to the steady-state probabilities, $p_{\rm st}(n_s)$, 
from Eq.~\ref{fpeStat}: 
\begin{equation}
\langle t_{r} \rangle (n_s)=\frac{\Delta t}{p_{\rm st}(n_s)}.\label{recurrencetime}
\end{equation} 
Here, $\Delta t$ is the time interval, corresponding to the integration time step between configurations. 
Then, the second derivative of the $N_c^{\rm th}$ passage time is given by 
\begin{eqnarray}
\frac{\partial^2\tau_{N_c}(n_s)}{\partial n_s^2}&=&\frac{1}{D_0}+
\frac{\beta \partial U(n_s)}{\partial n_s}\frac{\partial\tau(n_s)}{\partial n_s}+ \nonumber \\
&&f(N_c)\frac{\partial^2 \langle t_r\rangle(n_s)}{\partial n_s^2}.\label{nthtime2a}
\end{eqnarray} 
Using Eqs.~\ref{fpeStat} and~\ref{recurrencetime}, we can expand the last term as 
\begin{eqnarray}
\frac{\partial^2\langle t_r\rangle(n_s)}{\partial n_s^2}&=&f(N_c)\langle t_r\rangle(n_s) \Biggl\{ \frac{\beta\partial^2 U(n_s)}{\partial n_s^2} + \nonumber \\ 
&& \biggl[ \frac{\beta \partial U(n_s)}{\partial n_s} + \frac{j}{D_0p_{\rm st}(n_s)} \biggr] \times \nonumber \\ 
&& \biggl[\frac{\beta \partial U(n_s)}{\partial n_s} + \frac{2j}{D_0p_{\rm st}(n_s)} \biggl] \Biggl\}. \label{nthtime2b}
\end{eqnarray}
In the steepest descent approximation, the free energy around the top of the barrier can be written as  
\begin{equation}
U(n_s) \approx U(n_s^*) - \frac{c^2}{\beta} \left (n_s-n_s^*\right)^2 ,\label{energybarrier}
\end{equation} 
and the first derivative of the MPFT is  
\begin{equation}
\frac{\partial\tau(n_s)}{\partial n_s} = \frac{c}{j\sqrt{\pi}} e^{- c^2\left (n_s-n_s^*\right)^2 }.\label{mfptderivative}
\end{equation} 
Then, we insert these expressions into Eq.~\ref{nthtime2a} and set it to zero to obtain the position of the 
inflection point as a function of $N_c$:
\begin{eqnarray}
0&=&\frac{\partial^2\tau_{N_c}(n_s)}{\partial n_s^2}=\frac{1}{D_0}-2c^2\left (n_s-n_s^*\right)\frac{c}{j\sqrt{\pi}} e^{ -c^2\left (n_s-n_s^*\right)^2 } +\nonumber \\ 
&&\frac{f(N_c)\Delta t}{p_{\rm st}(n_s)}  \Biggl\{ -2c^2 +  \biggl [  \frac{j}{D_0p_{\rm st}(n_s)} - 2c^2\left (n_s-n_s^*\right)\biggr ] \times  \nonumber \\ 
&& \biggl [  \frac{2j}{D_0p_{\rm st}(n_s)} - 2c^2\left (n_s-n_s^*\right)\biggr ]  \Biggr \} .\label{nthtime2c}
\end{eqnarray} 
Next, we assume that the stationary probability distribution close to the top of the barrier can be approximated by the 
equilibrium distribution: 
\begin{eqnarray}
p_{\rm st}(n_s) &\approx &p_{\rm eq}(n_s) =  Z^{-1} e^{-\beta U(n_s)} \nonumber \\
&\approx&  Z^{-1} e^{-\beta U(n_s^*)+c^2(n_s-n_s^*)^2},\label{eqdist}
\end{eqnarray} 
where $Z=\int_a^b {\rm d}z e^{-\beta U(z)}$ is the normalization factor of a probability distribution defined on an interval $(a,b)$.  
In addition, we assume that the reaction rate is well approximated by classical nucleation theory, 
\begin{equation}
j \approx \frac{c}{\sqrt{\pi}}e^{-\beta U(n_s^*)},\label{nuclrateapprox}
\end{equation} 
and ignore all terms $\sim D_0^{-1}$. In this way, we obtain a quadratic equation for the position of the inflection 
point, 
\begin{equation}
\tilde{n}_s(N_c)=n_s^*+\frac{\sqrt{\pi}Z^{-1}}{2c^2\Delta t f(N_c)}\left \{ 1 \pm \sqrt{1+\frac{8 \left [c\Delta t f(N_c)\right ]^2}{\pi Z^{-2}}}\right \}. \label{position}
\end{equation} 

\bibstyle{revtex}

\end{document}